\documentclass[letterpaper,showpacs,twocolumn,eqsecnum,superscriptaddress,nofootinbib]{revtex4}
\usepackage[utf8]{inputenc}

\usepackage{amssymb,amsmath,amsthm,amsfonts,epsfig,times,natbib}
\usepackage{graphicx} 

\usepackage[usenames,dvipsnames]{color}
\usepackage{epstopdf}
\usepackage{amsmath,amssymb}
\usepackage{tensor}
\usepackage{mathtools}
\usepackage{amsbsy}
\usepackage{bm,url}
\usepackage{hyperref}

\begin{document}

\title{Dynamical $\ell$-boson stars: generic stability and evidence for non-spherical solutions}

\author{V\'ictor Jaramillo} 
\affiliation{Instituto de Ciencias Nucleares, Universidad Nacional
  Aut\'onoma de M\'exico, Circuito Exterior C.U., A.P. 70-543,
  M\'exico D.F. 04510, M\'exico}

\author{Nicolas Sanchis-Gual}
\affiliation{Centro de Astrof\'\i sica e Gravita\c c\~ao - CENTRA, Departamento de F\'\i sica,
Instituto Superior T\'ecnico - IST, Universidade de Lisboa - UL, Avenida
Rovisco Pais 1, 1049-001, Portugal}

\author{Juan Barranco}
\affiliation{Departamento de F\'isica, Divisi\'on de Ciencias e Ingenier\'ias,
Campus Le\'on, Universidad de Guanajuato, Le\'on 37150, M\'exico}

\author{Argelia Bernal}
\affiliation{Departamento de F\'isica, Divisi\'on de Ciencias e Ingenier\'ias,
Campus Le\'on, Universidad de Guanajuato, Le\'on 37150, M\'exico}

\author{Juan Carlos Degollado} 
\affiliation{Instituto de Ciencias F\'isicas, Universidad Nacional Aut\'onoma 
de M\'exico, Apdo. Postal 48-3, 62251, Cuernavaca, Morelos, M\'exico}

 \author{Carlos Herdeiro}
\affiliation{Departamento  de  Matem\'atica  da  Universidade  de  Aveiro and  Centre  for  Research  and  Development  in Mathematics  and  Applications  (CIDMA), Campus  de  Santiago,  3810-183  Aveiro,  Portugal}

\author{Dar\'{\i}o N\'u\~nez}
\affiliation{Instituto de Ciencias Nucleares, Universidad Nacional
  Aut\'onoma de M\'exico, Circuito Exterior C.U., A.P. 70-543,
  M\'exico D.F. 04510, M\'exico}


\date{April, 2020}

\begin{abstract}
$\ell$-boson stars are static, spherical, multi-field self-gravitating solitons. They are asymptotically flat, finite energy solutions of Einstein's gravity minimally coupled to an odd number of massive, complex scalar fields. A previous study assessed the stability of $\ell$-boson stars under spherical perturbations, finding that there are both stable and unstable branches of solutions, as for single-field boson stars ($\ell=0$). In this work we probe the stability of $\ell$-boson stars against non-spherical perturbations by performing numerical evolutions of the Einstein-Klein-Gordon system, with a 3D code. For the timescales explored, the $\ell$-boson stars belonging to the spherical stable branch do not exhibit measurable growing modes. 
We find, however, evidence of zero modes; that is, non-spherical perturbations that neither grow nor decay. This suggests the branching off towards a larger family of equilibrium solutions: we conjecture that $\ell$-boson stars are the enhanced isometry point of a larger family of static (and possibly stationary), non-spherical multi-field self-gravitating solitons.
\end{abstract}



\maketitle
\section{Introduction}
\label{sec:introduction}
Boson stars~\cite{Kaup68,Ruffini69} (see~\cite{Schunck:2003kk,Liebling:2012fv} for reviews) are remarkable gravitational solitons. These self-gravitating, localised energy lumps of a complex, massive scalar field have appealing theoretical properties. A key one is their dynamical stability. For spherical boson stars there is a stable branch of solutions against perturbations. Indeed, a variety of studies including linear perturbation theory \cite{1989NuPhB.315..477L,Gleiser:1988rq,Gleiser:1989a}, catastrophe theory \cite{Kusmartsev:1991pm} and 
numerical simulations \cite{Hawley2000, Seidel90, Guzman09, sanchis2017numerical, di2018dynamical} agree that boson stars are perturbatively stable, as long as the amplitude of the scalar field  is smaller 
than a critical value. When the latter is attained, boson stars acquire their maximum mass. 

Being dynamically stable legitimates inquiring about the possible (astro)physical role of boson stars. Albeit exotic, lacking undisputed observational evidence, boson stars have found important applications in strong gravity and astrophysics. For instance, boson stars provide a common model for a black hole mimicker~\cite{Mielke:2000mh,Torres2000,Guzman:2009zz,Olivares:2018abq}. Being dynamically tractable, one can then compare dynamical spacetime properties, such as waveforms of binary boson star systems, with those of black holes~\cite{Palenzuela:2006wp,Palenzuela:2007dm,bezares2017final}. This is particularly timely in view of the recently initiated gravitational-wave astronomy era~ \cite{Abbott:2016blz,TheLIGOScientific:2016qqj}, which provides data for both models to be compared with.

A second important application is in relation to a central mystery of contemporary science: the nature of dark matter. An increasing attention has been dedicated to models that consider dark matter as an ultra-light bosonic particle~\cite{Matos:1999et,Matos:2000ss,Hu:2000ke,Hui:2016ltb}. The bosonic nature allows this sort of dark matter to form coherent macroscopic excitations. 
In this context, bosons stars can model, in particular, the core of dark matter galactic halos \cite{Lee:1995af,Bernal:2009zy,UrenaLopez:2010ur,Schive:2014dra}.

In their original guise, the Einstein-Klein-Gordon (EKG) model contains a massive, free scalar field, and the solitonic solutions are called \textit{mini}-boson stars. A variety of generalisations ensued. Boson stars for scalar field theories with self-interactions have been reported, starting with the case of quartic self-interactions considered by Colpi \textit{et al.}~\cite{Colpi86}. Spacetime angular momentum was introduced for mini-boson stars in~\cite{Yoshida:1997qf,Schunck:1996he}, giving rise to stationary (but not static) self-gravitating solitons. A cousin model with a complex, massive vector (rather than scalar) field yields \textit{Proca stars}~\cite{brito2016proca}. These and other examples use single (complex) field models; however, multi-field boson stars have also been reported. One example is given by \textit{multi-state} boson stars \cite{Bernal:2009zy,UrenaLopez:2010ur,Li:2019mlk}. More  recently, multi-field boson stars with 
an arbitrary odd number, $N=2\ell +1$, $\ell\in \mathbb{N}_0$, of equal mass, uncoupled (except through gravity) complex scalar fields with harmonic time dependence were introduced~\cite{Alcubierre:2018ahf}; they are dubbed \textit{$\ell$-boson stars} and they will be the focus of this paper.

$\ell$-boson stars are described by spherically symmetric and static metrics. For $\ell=0$ they are simply the usual mini-boson
stars. For $\ell \geqslant 1$, the $2\ell +1$ scalar fields have an angular dependence given by the corresponding $2\ell +1$ spherical harmonics $Y^{\ell m}$. Then, if the radial dependence for all fields is the same, corresponding to choosing the amplitude of the spherical harmonics equal at all radial distances, static, spherical configurations are obtained, regardless of the energy-momentum tensor of each individual field being angular dependent. This is an example of \textit{symmetry non-inheritance}: the (spherical) spacetime and the (non-spherical) individual matter fields do not share spherical symmetry. The usual boson stars already have a version of symmetry non-inheritance: the (time oscillating) scalar field and the (static) spacetime do not share time-translation symmetry. Consistency requires only that the spacetime geometry and the total energy-momentum tensor share the same symmetries, not the individual matter fields.

Generic $\ell$-boson stars have been shown to exhibit similar properties to those of the standard $\ell=0$ stars. In particular, $\ell$- boson stars have a stable branch of solutions against   
\textit{spherical} perturbations  \cite{Alcubierre:2019qnh}. The main goal of this paper is to assess the stability of $\ell$-boson stars (in this branch) against generic, non-spherical perturbations. As we shall see, our analysis will show that, in this respect, generic $\ell$-boson star do not exactly mimic the $\ell=0$ case. Although no instabilities are observed, the analysis  provides a glimpse of a larger landscape of solutions, of which $\ell$-boson stars are just the enhanced symmetry point.

Departure from spherical symmetry is physically relevant. Firstly, spherical objects -- such as $\ell$-boson stars -- need to be stable against non-spherical perturbations, in order to be dynamically viable. Secondly, astrophysical bodies are not, typically, perfectly spherical, in particular due to angular momentum. So one must assess if some perturbations actually deform $\ell$-boson stars into acquiring new degrees of freedom. In this respect, it was recently proposed that multi-field boson stars, in the \textit{non-relativistic regime}, could have non-spherical stable configurations ~\cite{guzman2019gravitational}. This provides an extra motivation to inquire about the behaviour of relativistic $\ell$-boson stars under non-spherical perturbations Finally, assessing non-spherical configurations and perturbations often yields a richer phenomenology. As a fruitful example, it was recently found that spinning, single-field mini-boson stars are unstable against non-axisymmetric perturbations, either decaying into a non-rotating boson star or collapsing into a Kerr black hole \cite{bezares2017final, sanchis2019nonlinear}. By contrast, spinning Proca stars do not present instabilities under non-axisymmetric perturbations and furthermore, they can form dynamically~\cite{sanchis2019nonlinear}. This example shows how the study on non-spherical perturbations unveiled a new relevant dynamical property of boson stars. 

We shall investigate the behaviour of $\ell$-boson stars under non-spherical perturbations using fully non-linear numerical simulations of the corresponding Einstein-Klein-Gordon system. 
As initial data, we use configurations found in \cite{Alcubierre:2018ahf} which are then perturbed in two different ways.
The first type of perturbation tests the stability against non-axially symmetric perturbations, targeting potential bar-mode instabilities.
The second type of perturbation tests the stability against a relative change in the amplitude of the internal fields.
In none of the two cases measurable growing modes were found, either by perturbing the total mass density or by perturbing each of the constituent fields, as long the $\ell$-boson star belongs to the stable branch against spherical perturbations. By following the evolution of distortion parameters (defined below) we found, however, evidence for long-lived perturbations, which we interpret as zero modes. These modes, in turn, are interpreted as evidence for a larger family of equilibrium solutions.

Consider the Schwarzschild black hole of vacuum General Relativity. It has been shown to be mode stable in the renowned works of Regge and Wheeler~\cite{Regge57} and Zerilli~\cite{Zerilli70}. No gravitational perturbations grow. However, a perturbation that carries angular momentum yield not decay. The Schwarzschild solution migrates to a small angular momentum Kerr solution and oscillates around this new ground state. Similarly, a perturbation which electric charge will not decay and the spacetime will oscillate around a small charge Reissner-Nordstr\"om solution. These special perturbations are zero modes. Such modes are often found when a spacetime is unstable against some sort of perturbations, at the threshold between stable and unstable modes. An example occurs for the superradiant instability of the Kerr spacetime due to a massive bosonic field. The zero modes indicate the bifurcation of the Kerr family towards a new family of black holes with bosonic hair~\cite{Herdeiro:2014goa,Herdeiro:2016tmi}. But zero modes can also occur even if there is no instability, as in the Schwarzschild example, indicating, nonetheless, an enlarged family of solutions (Kerr or Reissner-Nordstr\"om), of which the initial spacetime (Schwarzschild) is a special case. Thus, one of the outcomes of our analysis is the conjecture that $\ell$-boson stars are the enhanced isometry point of a larger family of static (and possibly stationary), non-spherical multi-field self-gravitating solitons.

In the rest of this work we will focus
on configurations with  $\ell=1$.
Such $\ell$-boson stars 
are described by $N=3$ fields, with  $m=-1, 0,1$ respectively.
In order to follow the dynamics of the perturbed system a numerical code that solves the Einstein-$N$-Klein-Gordon system is required. We have used the \textsc{Einstein Toolkit} framework~\cite{EinsteinToolkit:web,loffler2012f, Zilhao:2013hia} with the
\textsc{Carpet} package~\cite{Schnetter:2003rb,CarpetCode:web} for
mesh-refinement capabilities to achieve our goal.

As a technical step we perform a Cauchy (3+1) decomposition on each scalar field that constitutes the star and solve the full Einstein-$N$-Klein-Gordon system.
This is done implementing an arrangement in the \textsc{Einstein Toolkit}, a {\it thorn}, to 
solve $N$ scalar fields using finite differences~\cite{sanchis2019nonlinear}.

This paper is organized as follows: Section~\ref{sec:id} 
addresses the construction of initial data to set up perturbed $\ell$-boson stars. Section~\ref{sec:diagnostics} describes the diagnostic tools
used to monitor the evolution and some aspects used to decide on whether instabilities are present. The numerical results are described
in Section~\ref{sec:time_evolution} and in Section~\ref{sec:discussion} our conclusions and final remarks are presented. In this work we use units where $G=1=c$.

\section{Initial Data}
\label{sec:id}

Following previous works on $\ell$-boson stars

\cite{Alcubierre:2018ahf,Alcubierre:2019qnh}, we consider a set of $N=2\ell+1$ complex scalar fields, with mass $\mu$ and no self-interaction within the Einstein theory of gravity, for which the energy-momentum tensor is given by:
\begin{equation}\label{eq:stressenergytotal}
T_{\alpha\beta}=\sum_{i=1}^N T^{(i)}_{\alpha\beta}\; ,
\end{equation}
where the index $i$ labels each field and the stress-energy-momentum for each field is given by
\begin{eqnarray}
\label{eq:stressenergy1f}
T^{(i)}_{\alpha\beta}&=&\left(\nabla_{\alpha}{\Phi_i}\,\nabla_{\beta}{\Phi^*_i} +
\nabla_{\beta}{\Phi_i}\,\nabla_{\alpha}{\Phi^*_i}\right) \nonumber \\ && +
g_{\alpha\beta}\left(
\nabla_{\sigma}{\Phi_i}\nabla^{\sigma}{\Phi^*_i}
+ \frac{1}{2}\mu^2 |\Phi_i|^2\right) \; . 
\end{eqnarray}
Complex conjugation is denoted by `*'. Following~\cite{Olabarrieta:2007di,Alcubierre:2018ahf} we propose a set of scalar fields of the form
\begin{equation}\label{eq:ansatzYlm}
\Phi^{(i)}(t,r,\vartheta,\varphi) = \psi_{\ell}(r,t) Y^{\ell m}(\vartheta,\varphi) \ ,
\end{equation}
where the angular momentum number $\ell$ is fixed, and $m$, which plays the role of index $i$ in equation (\ref{eq:ansatzYlm}), takes the values   $m = -\ell,-\ell+1,\ldots,\ell$ (hence the total number of fields needed for a fixed value of $\ell$ will be $2\ell+1$),  $Y^{\ell m}$ are the spherical harmonics defined over the unitary 2D-sphere. Then we assume that the amplitudes $\psi_{\ell}(r,t)$
are the {\it same} for all $m$. 
It was shown in \cite{Alcubierre:2018ahf} that if the $N$ fields have all the same amplitude $\psi_{\ell}$,
the stress-energy tensor \eqref{eq:stressenergytotal} has spherical symmetry regardless if the fields have angular dependence. See also \cite{Olabarrieta:2007di} and for a detailed discussion on the procedure, see \cite{Carvente:2019gkd}.

Assuming the harmonic time dependence 
\begin{equation}
\psi_\ell(r,t) = \phi_{\ell}(r) e^{-i \omega t} \ , 
\end{equation}
where $\phi_{\ell}(r)$ and the frequency $\omega$ are both real-valued, the stress-energy tensor becomes time independent. 
Under these assumptions it is possible to find self gravitating static, spherically symmetric equilibrium configurations by solving the EKG system of equations. Those configurations are parametrized by the angular momentum number $\ell$, hence the name, $\ell$-boson stars.

In order to obtain initial data suitable for numerical evolution, we construct equilibrium $\ell$-boson stars, to be subsequently perturbed. Considering a spherically symmetric spacetime with a line element given by:
\begin{equation}\label{eq:metric3p1}
ds^2 = - \alpha(r)^2 dt^2 + A(r)dr^2 + r^2(d\vartheta^2 + \sin^2\vartheta d\varphi^2 )\ ,
\end{equation}
where $\alpha$ and $A$, are functions of $r$, and the assumptions mentioned above for $\psi_\ell$, the EKG system yields
\begin{eqnarray}\label{eq:bosonsphericalp}
\partial^2_r \phi_{\ell} &=& -  \partial_r \phi_{\ell} \left( \frac{2}{r} + \frac{\partial_r \alpha}{\alpha}
- \frac{\partial_r A}{2A} \right)\nonumber\\
&+& A \phi_{\ell} \left( \mu^2 + \frac{\ell(\ell+1)}{r^2} - \frac{\omega^2}{\alpha^2} \right) \: ,
\end{eqnarray}
\begin{eqnarray}\label{eq:bosonsphericalA}
\partial_r A &=& A \left\{ \frac{(1-A)}{r} + 4 \pi r A \left[ \frac{(\partial_r{\phi}_{\ell})^2}{A} \right. \right. \nonumber \\
&+& \left. \left. \phi^2_{\ell} \left( \mu^2 + \frac{\ell(\ell+1)}{r^2}
+ \frac{\omega^2}{\alpha^2} \right) \right] \right\} \: , 
\end{eqnarray}
\begin{eqnarray}\label{eq:bosonsphericalAlpha}
\partial_r \alpha &=& \alpha \left[ \frac{(A-1)}{r} + \frac{\partial_r A}{2A} \right. \nonumber \\
&-& \left. 4 \pi r A {\phi^2_{\ell}} \left( \mu^2 + \frac{\ell(\ell+1)}{r^2} \right) \right] \: .
\end{eqnarray}

By studying the Klein-Gordon equation in the vicinity of $r=0$ one finds that the scalar field behaves as $\phi \sim \phi_0 r^{\ell}$ in that region. For a fixed value of the angular momentum number $\ell$, a given value of the parameter $\phi_0$, and the boundary condition at infinity requesting that $\phi_{\ell}$ decays exponentially, the system of equations (\ref{eq:bosonsphericalp}-\ref{eq:bosonsphericalAlpha}) becomes a nonlinear eigenvalue problem for the frequency $\omega$. 
We solve this set of equations in a finite size grid by means of a shooting method using the frequency $\omega$ as the shooting parameter.
For numerical purposes
we take the mass parameter $\mu = 1$.

\begin{figure}
\includegraphics[width=0.5\textwidth]{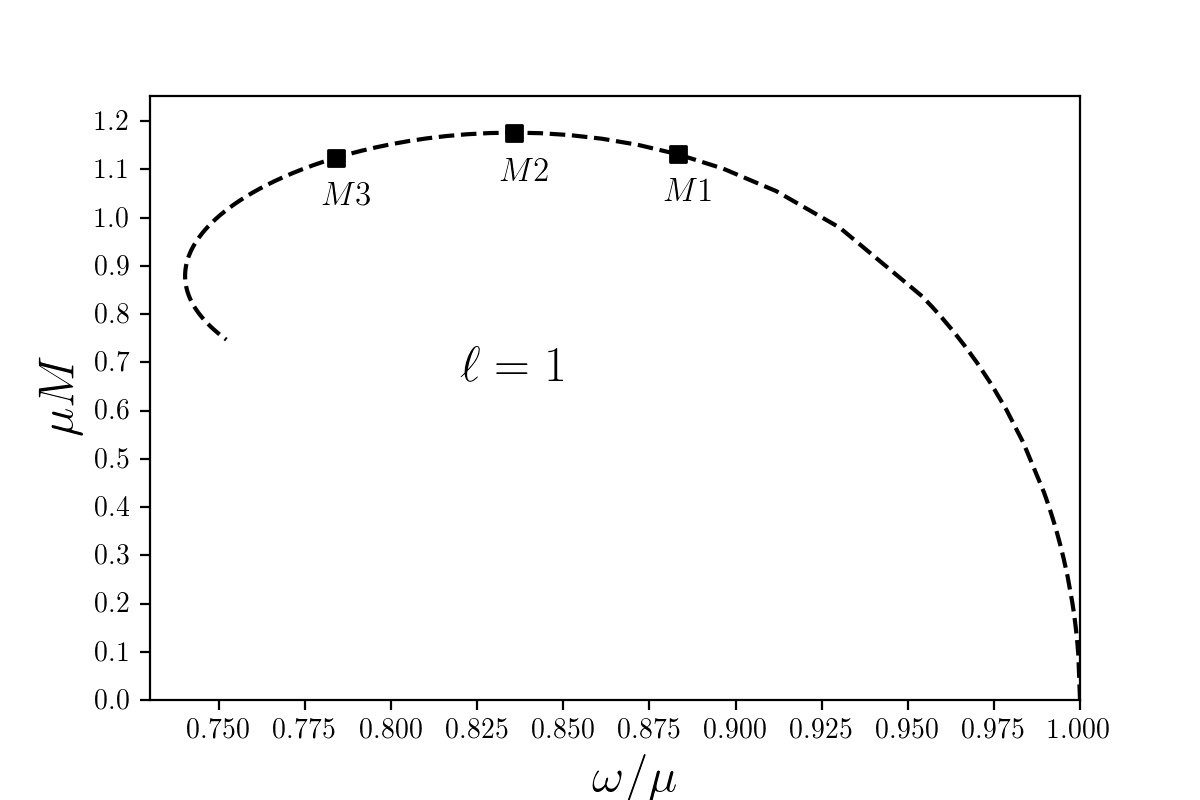}
\caption{ADM mass vs. frequency for static $\ell$-boson stars. The properties of models $M1$, $M2$ and $M3$ are listed in Table \ref{tab:models}.} 
\label{fig:mass_vs_frequency}
\end{figure}

Fig.~\ref{fig:mass_vs_frequency} shows a plot of the Arnowitt-Deser-Misner (ADM) mass $M$ versus the frequency $\omega$ for the $\ell$-boson stars. In Ref.~\cite{Alcubierre:2018ahf} it was shown that $\ell$-boson stars with $\ell > 0$ have 
similar properties to those of single-field mini-boson stars, $i.e.$ the $\ell=0$ case.
For instance, given a value of $\ell$, the mass $M$  of the equilibrium configurations as a function of $\omega$ has a maximum, which gets larger as $\ell$ increases, yielding more compact stars.
Furthermore, as in the case of $0$-boson stars, the maximum value of the mass separates the space of solutions into two branches. These branches correspond to stable and unstable configurations against spherical perturbations, as shown in \cite{Alcubierre:2019qnh}.

As mentioned above, the hypothesis that all the fields must have the same amplitude is essential to keep the spherical symmetry of the configuration. If one wants to consider different amplitudes of each constituent field, the assumption of spherical symmetry has to be relaxed. However, hitherto there has been no evidence that the resulting states may be equilibrium solutions of the Einstein-$N$-Klein-Gordon system.
In this work we will show that
deviations from spherical symmetry may indeed lead to new equilibrium solutions. 

To proceed further with our non-spherical analysis we transform the solutions of the previous system of equations to Cartesian coordinates, $x^{\mu}=(t,r,\vartheta,\varphi)$ $\rightarrow$
$x^{\mu}=(t,x,y,z)$.
Then we perform
a full non-linear numerical evolution of the perturbed stationary solutions.

\section{Diagnostics}
\label{sec:diagnostics}

In order to test the stability of the static solutions 
we perform two different types of perturbations: 

(i) The first type consists in perturbing
the energy density of the star given by $\rho = n^{\alpha}n^{\beta}T_{\alpha\beta}$, where $n^{\alpha}$ is the four velocity of Eulerian observers in the 3+1 space time decomposition.
The perturbed energy density 
is obtained by adding a non spherically symmetric small amplitude term to the homogeneous density in the following way
\cite{Saijo:2000qt}:
\begin{equation}\label{eq:rho_pert}
\rho=\rho_0\left[1+\kappa\left(\frac{x^2-y^2}{R_{99}^2}\right)\right]
\end{equation}
where $\rho_0$ is the energy density of the equilibrium configuration, obtained from the solution of~Eqs.~(\ref{eq:bosonsphericalp})-(\ref{eq:bosonsphericalAlpha}),
and $R_{99}$ is the radius enclosing 99\% of the configuration's mass. In our simulations we choose $\kappa = 0.1$.
This type of perturbations
could trigger
a potential bar-mode 
instability because
it only affects the $I_{xx}$ and $I_{yy}$ components of the quadrupole moment defined as
\begin{equation}
I_{xx} = \int \rho(y^2+z^2)\, dV \ , \qquad 
I_{yy} = \int \rho(x^2+z^2)\, dV \  . \quad
\end{equation}
Since the value of $\kappa$ is small $\kappa\ll1$, this perturbation can be considered
linear, initially; more importantly, it breaks the spherical symmetry of the original solution.

(ii) The second type of perturbations consist in varying separately the amplitude 
of each field. With these perturbations 
it is possible to
study the stability of the stars against variations on each mode $m$ and break the spherical symmetry.
We choose the following form 
\begin{eqnarray}\label{eq:perturbation}
    \phi_{\ell,m} = (1+\epsilon)\phi_{\ell},
\end{eqnarray}
where $\phi_{\ell}$ is the unperturbed solution 
of the system of Eqs.~(\ref{eq:bosonsphericalp})-(\ref{eq:bosonsphericalAlpha}).
This perturbation introduces an additional constraint violation, besides the well known numerical error, but its magnitude is controlled by choosing a small $\epsilon$, which, in general, depends on the $\ell,m$-mode. Note that if $\epsilon$ is the same for all $m$, the perturbation is spherical.

In order to assess the stability properties of the stars
during the numerical simulation, we monitor
the mass of the star, its angular momentum, and its density.
We also follow the change in the quadrupole moment of the star, as shown 
below. Following the technique described in \cite{Saijo:2000qt, Shibata:2010wz} to examine the stability of rotating neutron stars, we monitor the behaviour of the distortion parameter defined as
\begin{eqnarray}
\eta_z := \frac{I_{xx}-I_{yy}}{I_{xx}+I_{yy}}\; ,
\end{eqnarray}
which is a good measure of the magnitude of the bar-mode instability for perturbation (i).
This parameter has been used to study the stability of rapidly, differentially rotating
stars~\cite{Saijo:2000qt}.
It has been observed that when the star is dynamically unstable, $\eta_z$ grows exponentially up to a maximum value; then the maximum value of $\eta_z$ remains constant on dynamical
timescales. 
For stable stars, on the other hand, the maximum initial value of $\eta_z$ remains constant throughout the evolution.
Thus the monitoring of $\eta_z$ provides a good tool to determine the properties of the star against bar-mode perturbations.
In this work we also use 
\begin{eqnarray}
\eta_y := \frac{I_{xx}-I_{zz}}{I_{xx}+I_{zz}}\; ,
\end{eqnarray}
as a measure of the deformation of the star.

As further diagonostics, the maximum
of the density and the lapse function are used
to determine whether the configuration disperses or is undergoing a collapse.  We have used the thorn \textsc{AHFinder}~\cite{diener2003new} to follow the formation of an apparent horizon (AH) during the evolution. We have also computed the Hamiltonian 
constraint \cite{Gourgoulhon2012} to check the fourth order convergence of the implementation -  
see the Appendix for details on this procedure.

\section{Time evolution and numerical results}
\label{sec:time_evolution}

In this section we present the results from dynamical
spacetime simulations 
from the perturbed $\ell$-boson stars. We have compared the evolution of the equilibrium $\ell$-boson stars with the perturbed stars.

We numerically integrate the EKG system
using fourth-order spatial discretization within the \textsc{Einstein Toolkit} framework.
The \textsc{Einstein Toolkit}
solves the Einstein equations within the ADM 3+1 framework and evolves the spacetime using the Baumgarte-Shapiro-Shibata-Nakamura (BSSN) formulation of the Einstein equations \cite{Baumgarte:1998te} 
through the \textsc{McLachlan} thorn~\cite{brown2009turduckening,reisswig2011gravitational}
. All the evolutions were made using the 1+log time slicing condition for the lapse $\alpha$, 
and the {\it Gamma-driver} condition for the shift $\beta^{i}$ \cite{Alcubierre08a}.

We use the Method of Lines thorn to solve the equations in time by using
a fourth order Runge-Kutta scheme. The
equations for the scalar fields
are solved using a finite difference scheme of fourth order.
We also employ the mesh refinement capabilities provided by the \textsc{Carpet} arrangements.
The fixed mesh refinement grid hierarchy used consists of
nested cubes with 3 levels of refinement. 
The finest is set in such a way that it covers the entire star.

We set the spatial resolution on the finest
level to $\{dx, dy, dz\}= 0.8$ (and the coarsest to $\{dx, dy,dz\}=3.2$) in order to fully capture the  properties of the star.
We follow the formation of an AH after the collapse of unstable stars.

More details on the resolution, as well as numerical convergence are given in the Appendix.

The three stationary configurations we chose to illustrate the general behaviour of the stars are represented with a square over the curve in Fig.~\ref{fig:mass_vs_frequency} denoted as ($M1$, $M2$, $M3$). Some of the  properties of these stars are summarised in Table I.

\begin{table}
		\begin{tabular}{c|c|c|c|c}  
			\hline 
			Model&$\ell$&$\omega/\mu$ & $\mu R_{99}$&$\mu M_{\text{ADM}}$\\
			\hline
			$M1$ & 1 & $0.882$ &13.45& 1.133\\
			$M2$ & 1&$0.836$ & 12.75&1.176 (maximum)\\
			$M3$ & 1&$0.783$ &7.53& 1.122\\
			\hline 
        \end{tabular}
	\caption {Frequency, radius and ADM mass for the configurations analysed.}\label{tab:models}
\end{table}

Both spherical and non-spherical perturbations to the stationary solutions 
are
induced by increasing or decreasing the amplitude of 
the different constituent fields, see Eq.~(\ref{eq:perturbation}). 
In our case of study, $\ell=1$ and thus, for each configuration $M1,M2,M3$ there are three fields $\phi_{\ell,m}$: $\lbrace$$\phi_{1,-1}$, $\phi_{1,0}$, $\phi_{1,1}$$\rbrace$.
We use the position of sub-index in the models ($M1_{m=-1,m=0,m=1}$) to label the mode (field) that is being perturbed.  We use $+$ or $-$ to ascribe an increase ($\epsilon > 0$) or decrease ($\epsilon<0$) of the amplitude,
we use $0$ to represent that no perturbation was introduced in that mode ($\epsilon=0$). In this way, for instance, $M1_{+0-}$ means that $M1$ has been perturbed in the following way: the first scalar field, $\phi_{1,-1}$, has been perturbed with $\epsilon>0$; the second field $\phi_{1,0}$ has not been perturbed ($\epsilon=0$), and the third field $\phi_{1,1}$ has been perturbed with $\epsilon<0$.
In summary, we have perturbed $M1,M2,M3$ in the following ways: perturbing all fields with the same amplitude as a test (spherical perturbation), introducing a non-axisymmetric bar-mode perturbation,
and finally, we 
perturbed each constituent field using different amplitudes.

\subsection{Spherical perturbation test}\label{sec:spherical}

\begin{table}
\begin{tabular}{l|c|c|c|c}
\hline
Model $M1$ ($\omega/\mu=0.882$)&\multicolumn{3}{| c |}{$\epsilon$}&Collapse\\
\hline
Run &$m=+1$&$m=0$&$m=-1$&\\
\hline
$M1_{000}$&0&0&$0$&  No\\
$M1_{+++}$&$+0.01$&$+0.01$&$+0.01$&No\\
$M1_{---}$&$-0.01$&$-0.01$&$-0.01$&No\\
\hline
Model $M2$ ($\omega/\mu=0.836$)&\multicolumn{3}{| c |}{$\epsilon$}&Collapse\\
\hline
$M2_{000}$&0&0&$0$&No\\
$M2_{+++}$&$+0.01$&$+0.01$&$+0.01$&Yes\\
$M2_{---}$&$-0.01$&$-0.01$&$-0.01$&No\\
\hline
Model $M3$ ($\omega/\mu=0.783$)&\multicolumn{3}{| c |}{$\epsilon$}&Collapse\\
\hline
$M3_{000}$&0&0&$0$&Yes\\
$M3_{+++}$&$+0.01$&$+0.01$&$+0.01$&Yes\\
$M3_{---}$&$-0.01$&$-0.01$&$-0.01$&No\\
\hline
\end{tabular}
\caption {List of simulations performed for the case where all fields 
are perturbed with the same amplitude (spherical perturbations). These cases are similar to the simulations performed in \cite{Alcubierre:2019qnh}.}\label{tab:spherical}
\end{table}

First, we perform numerical evolutions of the models listed in Table \ref{tab:models} with spherical perturbations. We induce perturbations in each field of a $\ell$-boson star with all the perturbations having the same amplitude. In this way we guarantee that the spherical symmetry is preserved. 
This type of perturbations is done in order to compare and validate our results with those found using a spherically symmetric 1D code reported in \cite{Alcubierre:2019qnh}. 
While perturbing the initial equilibrium configurations adding perturbations (with a positive or negative value for $\epsilon$) that preserve the spherical symmetry, we find that the configuration that was reported to be stable in Ref.~\cite{Alcubierre:2019qnh} (model $M1$) remains stable in the timescale we reach in the 3D simulations, run $M1_{000}$.
The values of the amplitude of the perturbations for these perturbed configurations are reported in Table \ref{tab:spherical}
as $M1_{+++}$ and $M1_{---}$, in which we perturb each field adding or subtracting $|\epsilon|=0.01$ to each mode. 

Our results are also consistent with models of $\ell$-boson stars that are unstable in spherical symmetry. 
According to the results in \cite{Alcubierre:2019qnh}, the configuration $M3_{000}$ is unstable in the 1D simulations. 
When we perturb the amplitudes of the fields adding (run $M3_{+++}$) or subtracting (run $M3_{---}$) the same amount, the configuration collapses or migrates to the stable branch respectively, as described in Table \ref{tab:spherical}.
We monitor the behaviour of the metric coefficients during the evolution, and, in particular, 
we use the lapse and the formation of an AH as an indicator of the collapse of the star and the formation of a black hole.

The model $M2$ deserves special mention since it corresponds to 

the critical solution:
the star with maximum mass. 
We found that perturbations increasing the amplitude of the field ($\epsilon>0$, run $M2_{+++}$)  make the star collapse whereas perturbations that decrease the amplitude ($\epsilon<0$, run $M2_{---}$) drive the configuration to a new stable state as described in Table \ref{tab:spherical}. These results are consistent with the results reported in \cite{Alcubierre:2019qnh} for perturbations that increase of decrease the mass of the star.

\subsection{Non-spherical perturbation: perturbing the energy density}

In order to determine whether $\ell$-boson stars develop a bar mode instability, 
we took as initial data a stationary  model and modified the energy density in accordance with eq. 
\eqref{eq:rho_pert}.
We have performed this analysis for configurations with $\ell=0$ and $\ell=1$, both stable against spherical perturbations. In the case $\ell=0$ we have taken the equilibrium configuration corresponding to $\omega/\mu=0.937$ and for $\ell=1$ the configuration with $\omega/\mu=0.882$, $M1$ in Table.~\ref{tab:models}.
By choosing $\kappa=0.01$, the momenta of inertia
$I_{xx}$ and $I_{yy}$ change by less than 0.5\% with respect to the equilibrium solution,
hence we consider that the induced initial perturbation is small.
Then we evolve the perturbed system via the Einstein-$N$-Klein-Gordon equations and monitor the behaviour of $\eta_z$.

\begin{figure}
\includegraphics[width=0.5\textwidth, height=0.27\textheight]{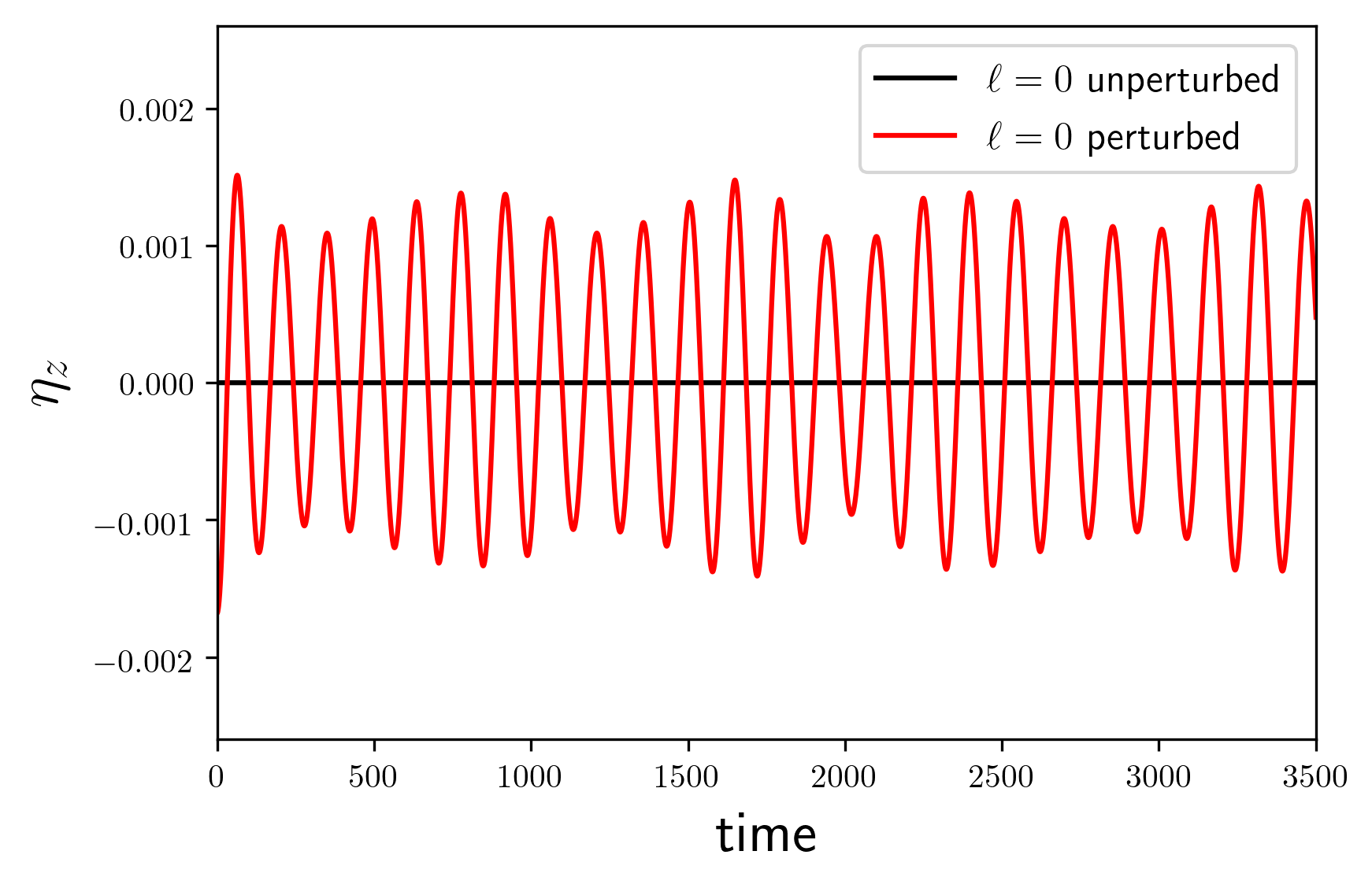}
\includegraphics[width=0.5\textwidth, height=0.27\textheight]{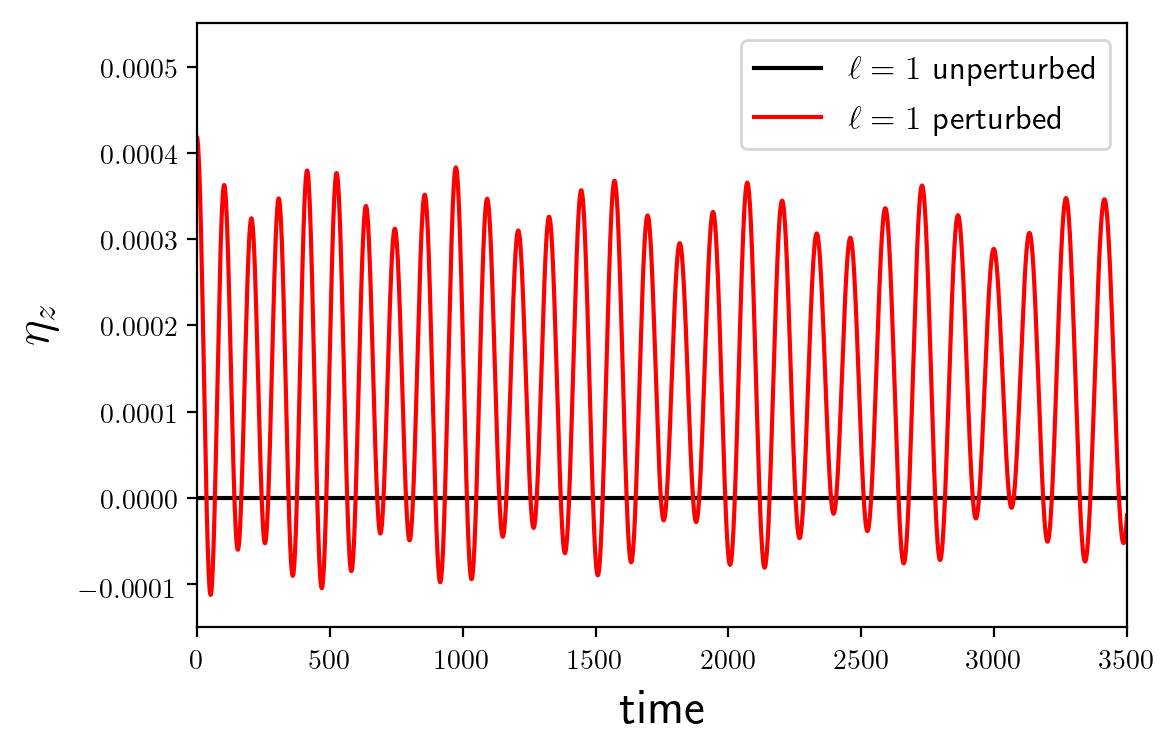}
\caption{Evolution of $\eta_z$ as a function of time for unperturbed (black solid line) and perturbed density as defined by Eq.~(\ref{eq:rho_pert}) (red solid line) with $\kappa=0.01$. Top panel: the unperturbed configuration is a single-field boson star ($\ell=0$). Bottom panel: the unperturbed configuration is a multi-field boson star ($\ell=1$). In neither perturbed case has a bar instability been observed. Instead, a long lived departure from spherical symmetry occurs for $\ell=1$, but not for $\ell=0$, as long as $\eta_z \ne 0$. In contrast, the unperturbed configurations preserve spherical symmetry as 
$\eta_z$ oscillates around zero in both cases.}
\label{fig:bar}
\end{figure}

In Fig.~\ref{fig:bar} we show $\eta_z$ as a function of time for perturbed and unperturbed configurations for $\ell=0$ (top panel) and $\ell=1$ (bottom panel). 
For $\ell=0$, the distortion $\eta_z$ oscillates around zero for the perturbed case, indicating 
the star maintains, essentially, the spherical symmetry.

On the other hand,  for $\ell=1$ in the case where the perturbation was included, the 
initial perturbation induces a small deviation from spherical symmetry therefore $\eta_z$ acquires a nontrivial value during the evolution.  
This non-zero value of $\eta_z$ indicates that the shape of the star deviates from spherical and becomes oblate. 

During the evolution time considered ($t\sim 3500$)  we did not find any signal of a bar-mode instability for the models considered: no exponential growth in $\eta_z$ was measured.
Most importantly $\eta_z$ does not grow as it 
happens for unstable stars \cite{sanchis2019nonlinear}.
This change in the shape of the $\ell=1$ configuration, illustrated by $\eta_z \ne 0$, is compared with the case where $M1$ is not perturbed (black solid line).
For the unperturbed case, $\eta_z$ simply oscillates 
around zero. 
The conclusion, therefore, is that the perturbed 
configuration lingers, neither collapsing nor dissipating, thus showing a
non-spherical distribution that is either stable or long-lived, without signs of instability. 
It is worth emphasising the key difference with the $\ell=0$ case, for which the evolution oscillates around a spherical distribution, in agreement with the fact that such a distribution is the only equilibrium configuration.

We found that after some time, the stars acquire a small linear momentum due to the numerical error and thus the deviation parameters can not be obtained accurately.
Once this becomes noticeable we stop the evolution. 

\subsection{Non-spherical perturbation: perturbing the amplitude of each mode}
\begin{table}
\begin{tabular}{l|c|c|c|c}
\hline
Model $M1$ ($\omega/\mu=0.882$)&\multicolumn{3}{| c |}{$\epsilon$}&Collapse\\
\hline
Run &$m=+1$&$m=0$&$m=-1$&\\
\hline
$M1_{+00}$&$+0.01$&0&$0$&No\\
$M1_{-00}$&$-0.01$&$0$&0&No\\
$M1_{0+0}$&0&$+0.01$&0&No\\
$M1_{0-0}$&$0$&$-0.01$&0&No\\
$M1_{00+}$&$0$&0&$+0.01$&No\\
$M1_{00-}$&0&0&$-0.01$&No\\
$M1_{++0}$&+0.01&$+0.01$&$0$&No\\
$M1_{+-+}$&$+0.01$&$-0.01$&$+0.01$&No\\
$M1_{+0+}$&$+0.01$&$0$&$+0.01$&No\\
$M1_{-0+}$&$-0.01$&$0$&$+0.01$&No\\
$M1_{0(-)0}$&$0$&$-0.1$&0&No\\
\hline
\end{tabular}
\caption {List of simulations performed 
for model $M1$ under the second type of
perturbations. The parenthesis indicate a larger amplitude on the perturbations. Notice that gravitational collapse was not observed in any of the simulations.}\label{tab:model1}
\end{table}

In this section, we describe the evolutions we have performed implementing non-spherical perturbation by varying the amplitude of each field of the $\ell$-boson star. 

\subsubsection{Non-spherical perturbation of $M1$}
\begin{figure}
\includegraphics[width=0.5\textwidth, height=0.27\textheight]{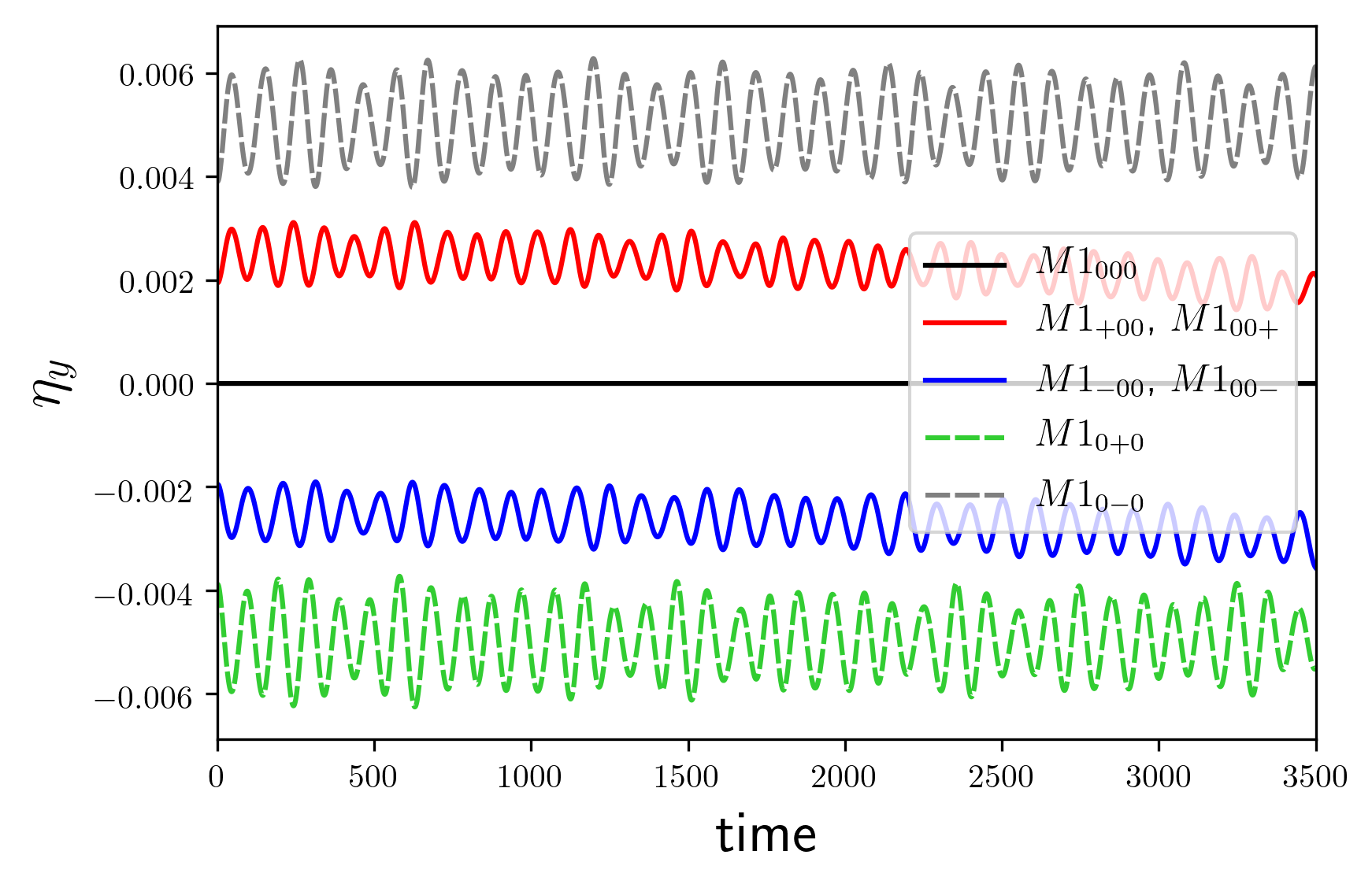}
\caption{Distortion parameter $\eta_y$, for model $M1$ for runs shown in Table \eqref{tab:model1} . Departure from spherical symmetry is shown for those configurations that have 
been perturbed differently
in all the 
three fields $\phi_{1,-1}$, $\phi_{1,0}$ and $\phi_{1,1}$. On the contrary, the configuration $M1_{000}$ has the same perturbation for all the fields (spherical perturbation), and it shows $\eta_y=0$ at all times.} 
\label{fig:distortion_y}
\end{figure}
In order to illustrate the procedure to perturb the star, let us consider first model $M1$. 
Different perturbations have been applied to $M1$ and all them are summarised in Table \ref{tab:model1}. Subscripts
indicate which fields have been perturbed and if the amplitude is increased by $\epsilon>0$ (subscript $+$), decreased by $\epsilon<0$ (subscript $-$) or it has been left without perturbation $\epsilon=0$ (subscript $0$), as described before.

All our evolutions show that $M1$ remains stable without collapsing (none AH was found), independently of the perturbation.
Thus, from the results summarised in Table \ref{tab:model1} we can conclude that the configurations in the stable branch (against spherical perturbations), that is, to the right of configuration $M2$ in Fig.~\ref{fig:mass_vs_frequency}, are also stable under non-spherical perturbations, against collapse.

Let us now turn to another result that can be extracted by studying the distortion parameters $\eta_z$ and $\eta_y$. For spherical configurations and for those configurations that are spherically perturbed, these are zero. On the other hand, for  non-spherical perturbations
a small deviation from spherical symmetry is induced. In other words, non-trivial values of $\eta_y$ are obtained throughout the evolution of $M1$. This behaviour of $\eta_y$  as a function of time is shown in Fig.~\ref{fig:distortion_y}. Indeed, non-zero values of $\eta_y$
for the evolution of    
$M1_{+00},M1_{00,+},M1_{-00},M1_{00-},M1_{0+0}$, and $M1_{0-0}$, are obtained. No instability is observed, but the deformation does not die off either. These long-lived deformed configurations, arising as dynamical solutions of 
the Einstein-$N$-Klein-Gordon with three fields with different $m$ are not spherically symmetric. The corresponding perturbations appear to be zero modes, suggesting a larger family of solutions.

As expected, we observe that the equilibrium configuration $M1_{000}$ has $\eta_y=0$ at all times of the evolution. 
Besides, we have found that the behaviour of $\eta_y$ and $\eta_z$ during the evolution is the same for perturbations in the modes $m=1$ and $m=-1$  
with the same values of $\epsilon$. This suggests that the resulting configurations are axially symmetric.   

\begin{figure}
\includegraphics[width=0.5\textwidth,height=0.27\textheight]{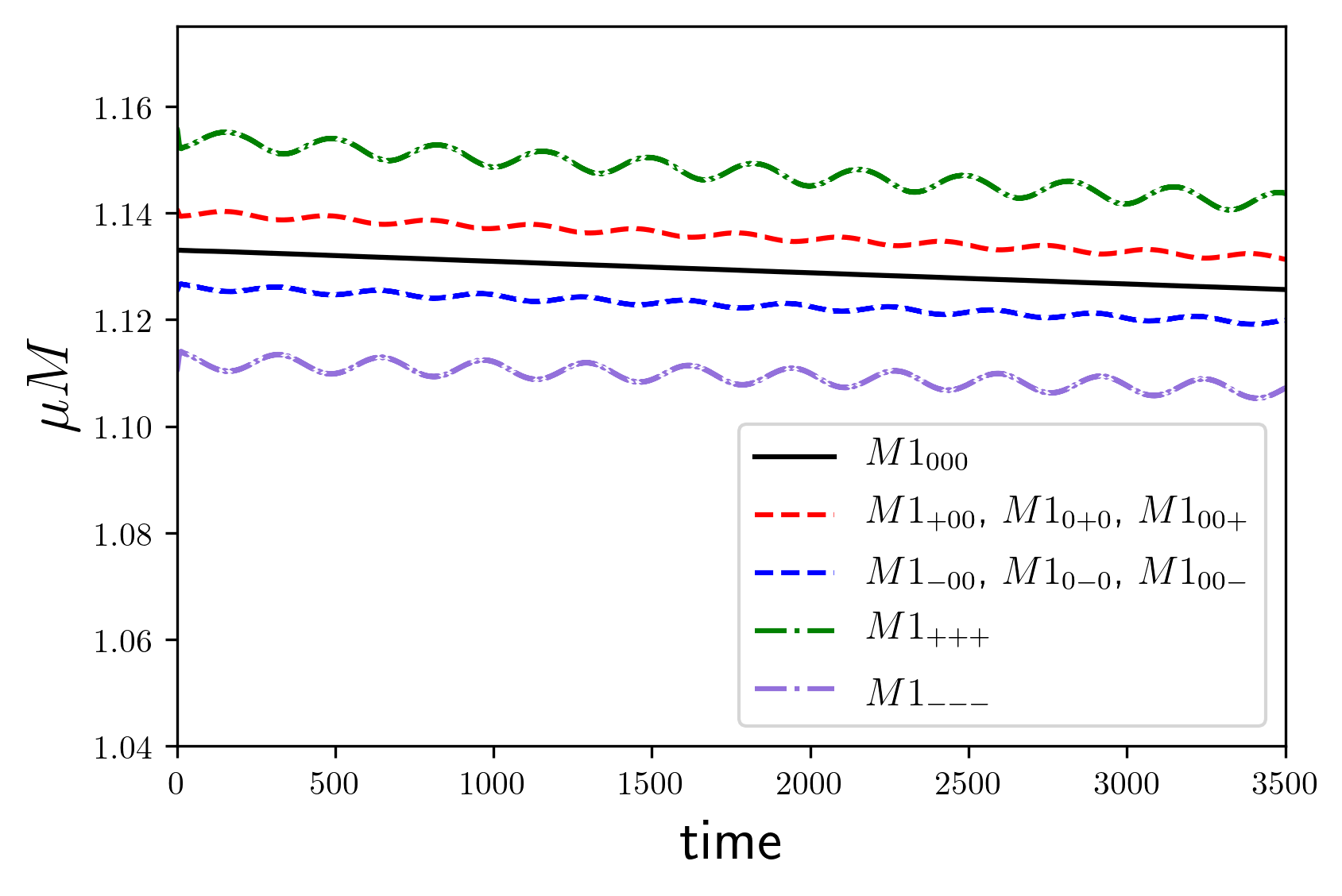}
\caption{Evolution of the mass of the model $M1$ subjected to spherical and non-spherical perturbations listed in Tables \ref{tab:spherical} and \ref{tab:model1} respectively.
The mass, as expected, is increased for those perturbations with $\epsilon>0$ and decreases when $\epsilon<0$.} 
\label{fig:mass_perturbed}
\end{figure}

Finally, we report in Fig.~\ref{fig:mass_perturbed} the time evolution of the total mass of $M1$ under  different 
spherical and non-spherical perturbations.
As expected, the mass of the perturbed configurations decreases or increases when $\epsilon<0$ or $\epsilon>0$.
Notice, however, the change in mass is the same whether we perturb the $m=-1$ or $m=1$ modes, for the same the sign of $\epsilon$. This fact supports the assertion that the resulting configurations are axially symmetric.

The total mass of the models decreases with time,
showing a small drift, even for the unperturbed solution. We have checked that this drift is due to the numerical error, since it is reduced when the grid resolution is increased (see Appendix A).

Fig.~\ref{fig:density} displays a series of snapshots of projections in the planes $xy$ and $xz$ of the energy density for the run $M1_{0(-)0}$
(With a large amplitude in the perturbation in the mode $m=0$)
. 
The initial perturbation is introduced in the mode $m=0$, decreasing 
the mass of the star and inducing a small deformation. 
Notice that we have taken the largest perturbation presented in this section (model  $M1_{0(-)0}$), so that the deformation can be appreciated in these projections of the energy density. At later times the system evolves and settles down into a configuration without collapsing or exploding. 
Although the configuration looks almost spherical, the value of $\eta_y$ at late times is slightly different from zero ($\eta_y\sim 0.05)$. We have evolved this configuration for $t\sim 10000$ and remains in the same state not showing any signs of instability, or returning to a $\ell$-boson star.   

\begin{figure}
\includegraphics[width=0.45\textwidth]{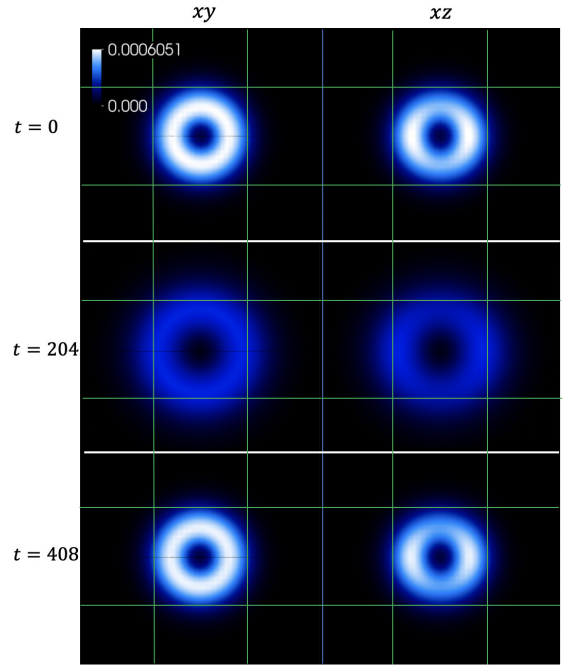}
\caption{Three snapshots of the projection of the rest mass density in two planes. In the second snapshot the star expands and thus the  maximum value of the density decreases.
In the third snapshot the star returns to its original state.
This repetitive behaviour is present during all the evolution time. The mesh represent a box with sides $\frac{1}{2}R_{99}$ of the unperturbed star.} 
\label{fig:density}
\end{figure}

At this point it is important to mention that perturbations in the mode $m = 0$ 
do not modify the value of the total angular momentum, while perturbations 
in the modes $m=1$ or $m=-1$
do. Specifically, $M1_{-00}$ and $M1_{00+}$ ($M1_{+00}$ and $M1_{00-}$) have a positive (negative), non-trivial and constant value of total angular momentum. As we could expect, runs like $M1_{+0+}$ have zero angular momentum. This particular result was also obtained for the perturbations of the models that will be presented below.

\subsubsection{Non-spherical perturbation of $M2$ and $M3$}
The results of the previous section indicate that those configurations ($M1$) that are stable under spherical perturbations  do not show non-spherical growing modes.
Furthermore, perturbations to the fields $\phi_{1,-1}$ and $\phi_{1,1}$ applied to those configurations provide evidence for zero modes, producing new equilibrium configurations that are dynamically stable, and have small departures from the
spherical configurations.

Now we are interested in studying non-spherical perturbations on configurations that might 
undergo gravitational collapse to a black hole. Those configurations  are $M2$ and $M3$. In particular $M2$, as we have mentioned, corresponds to the critical configuration with maximum ADM mass. Configuration $M2$, as is shown in Section \ref{sec:spherical} and reported in the literature \cite{Alcubierre:2019qnh}, divides stable from unstable configurations. The latter are those configurations that can collapse into a black hole.  We now go one step further and study them under non-spherical perturbations.  
\begin{table}
\begin{tabular}{l|c|c|c|c}
\hline
Model $M2$ ($\omega/\mu=0.836$)&\multicolumn{3}{| c |}{$\epsilon$}&Collapse\\
\hline
Run&$m=+1$&$m=0$&$m=-1$&\\
\hline
$M2_{+00}$&$+0.01$&0&$0$&Yes\\
$M2_{-00}$&$-0.01$&$0$&0&No\\
$M2_{0+0}$&0&$+0.01$&0&Yes\\
$M2_{0-0}$&$0$&$-0.01$&0&No\\
$M2_{00+}$&$0$&0&+0.01&Yes\\
$M2_{00-}$&0&0&$-0.01$&No\\
$M2_{++0}$&$+0.01$&+0.01&$0$&Yes\\
$M2_{+-+}$&$+0.01$&$-0.01$&$+0.01$&Yes\\
$M2_{+0+}$&$+0.01$&$0$&$+0.01$&Yes\\
$M2_{-0+}$&$-0.01$&$0$&$+0.01$&No\\
\hline
\end{tabular}
\caption{List of simulations performed from the model $M2$ under non-spherical perturbations. Those configurations that increased the total mass by the addition of the perturbation did collapse to a black hole. Configurations that did not change the total mass or did not decrease the total mass of the configuration did not collapse.}\label{tab:model2}
\end{table}

The list of perturbation applied to each field of $M2$ is summarised in Table \ref{tab:model2} and the result of the evolution is reported in the fifth column of the same Table. The results confirm that $M2$ is the configuration that separates stable from unstable configurations. Indeed, as it can be observe in the results of Table \ref{tab:model2}, all those configurations which have perturbations that increased the total mass of the configuration undergo a collapse, while those configurations which have been perturbed and reduced the total mass of the configuration did not collapse to a black hole. In this respect, run $M2_{-0+}$ is of special interest. The perturbation did not change the total mass of the configuration, and the result of their evolution is that it did not collapse.

Finally, we have considered non-spherical perturbations of model $M3$. The results of the evolution of the different models studied are summarised in Table \ref{tab:model3}. This configuration is on the so called unstable branch. The results mimic, to some extent, those observed for $M2$. Perturbations that reduced the total mass of the configuration led to a migration to the stable branch. On the other hand, those that increased the mass of the configuration let to a collapse into a black hole. But a key difference is seen for $M3_{-0+}$. This  perturbation did not change the total mass of the configuration, and contrary to $M2_{-0+}$, it did collapse to a black hole.
This result, combined with the spherical perturbations mentioned in \ref{sec:spherical}, further confirms the special status of the maximum mass configuration $M2$: it marks the threshold of unstable configurations of $\ell$-boson stars, for both spherical and non-spherical perturbations.
\begin{table}
\begin{tabular}{l|c|c|c|c}
\hline
Model $M3$ ($\omega/\mu=0.783$)&\multicolumn{3}{| c |}{$\epsilon$}&Collapse\\
\hline
Run&$m=+1$&$m=0$&$m=-1$&\\
\hline
$M3_{+00}$&$+0.01$&0&$0$&Yes\\
$M3_{-00}$&$-0.01$&$0$&0&No\\
$M3_{0+0}$&0&$+0.01$&0&Yes\\
$M3_{0-0}$&$0$&$-0.01$&0&No\\
$M3_{00+}$&$0$&0&$+0.01$&Yes\\
$M3_{00-}$&0&0&$-0.01$&No\\
$M3_{++0}$&$+0.01$&$+0.01$&$0$&Yes\\
$M3_{+-+}$&$+0.01$&$-0.01$&$+0.01$&Yes\\
$M3_{+0+}$&$+0.01$&$0$&$+0.01$&Yes\\
$M3_{-0+}$&$-0.01$&$0$&$+0.01$&Yes\\
\hline
\end{tabular}
\caption {List of simulations performed 
form the model $M3$ under non-spherical perturbations. 
Those configurations that increased the total mass by the addition of the perturbation did collapse to a black hole. Configurations that decreased the total mass of the initial configuration, 
migrated to the stable branch. The run $M3_{-0+}$ that did not change the total mass did collapse to a black hole.
}\label{tab:model3}
\end{table}

\section{Discussion and Outlook}
\label{sec:discussion}
In this paper we performed dynamical simulations in the 
fully non-linear EKG model to investigate the stability of
$\ell$-boson stars. Unlike previous works we have considered non-spherical perturbations. 
An expected result is that those configurations known to be unstable under spherical perturbations, are also unstable under more general perturbations. The most interesting question, however, was if the configurations known to be stable under spherical perturbations would remain stable under more general ones. Here, our conclusions are two-fold. Firstly, no growing modes have been measured in our simulations. In this sense $\ell$-boson stars are stable against non-spherical perturbations. However, when deformed away from sphericity, $\ell$-boson stars do not return to a spherical state. They appear to oscillate around a new (slightly) non spherical state. We take this as evidence that new, multi-field, equilibrium configurations of the Einstein-$N$-Klein Gordon system exist, which are non-spherical. This conjecture is our second conclusion.

If our conjecture is proven correct, the spherically symmetric $\ell$-boson stars are only an enhanced isometry point of a larger family of solutions of the Einstein-$N$-Klein Gordon. As discussed in the introduction, this is analogous to the Schwarzschild BH being the isometry enhancement point of the Kerr family. It is well known that the Kerr solution brings about qualitatively novel features with respect to the Schwarzschild solution. So it will be quite interesting to understand the novelties brought by the enlarged family of solutions that this work is suggesting.

The conjecture on the existence of these new non-spherical, multi-field configurations can be tested by solving the Einstein-$N$-Klein Gordon system for static or stationary (\textit{i.e.} spinning) configurations, without assuming spherical symmetry. Research in this direction is already ongoing.


\acknowledgments
We thank  Miguel Alcubierre, Alberto Diez, Miguel Meguevand, Eugen Radu and Olivier 
Sarbach, for stimulating discussions.
This work was supported in part by the CONACYT Network Project 
280908 ``Agujeros Negros y Ondas Gravitatorias", by DGAPA-UNAM through grants 
IN110218, IA103616, IN105920, by the European Union's Horizon 2020 research and innovation (RISE)
program H2020-MSCA-RISE-2017 Grant
No. FunFiCO-777740, by  the Center  for  Research  and  Development  in  Mathematics  and  Applications  (CIDMA)  through  the Portuguese Foundation for Science and Technology (FCT - Fundacao para a Ci\^encia e a Tecnologia), references UIDB/04106/2020 and UIDP/04106/2020, and by  the  projects  PTDC/FIS-OUT/28407/2017,  CERN/FIS-PAR/0027/2019 and UID/FIS/00099/2020 (CENTRA). 
The authors would like to acknowledge networking support by the COST Action CA16104. VJ acknowledge support from CONACYT.

\appendix
\section{Code validation}

For run $M1_{000}$ we report the time evolution of the mass and violations of the Hamiltonian constraint, with different resolutions, 
$\{dx,dy,dz\} =3.2$, $\{dx,dy,dz\} =\sqrt{2}\ 1.6$ and $\{dx,dy,dz\}= 1.6$, where $dx$, $dy$ and $dz$ are the sizes of the coarsest level of refinement

\begin{figure}
\includegraphics[width=0.5\textwidth]{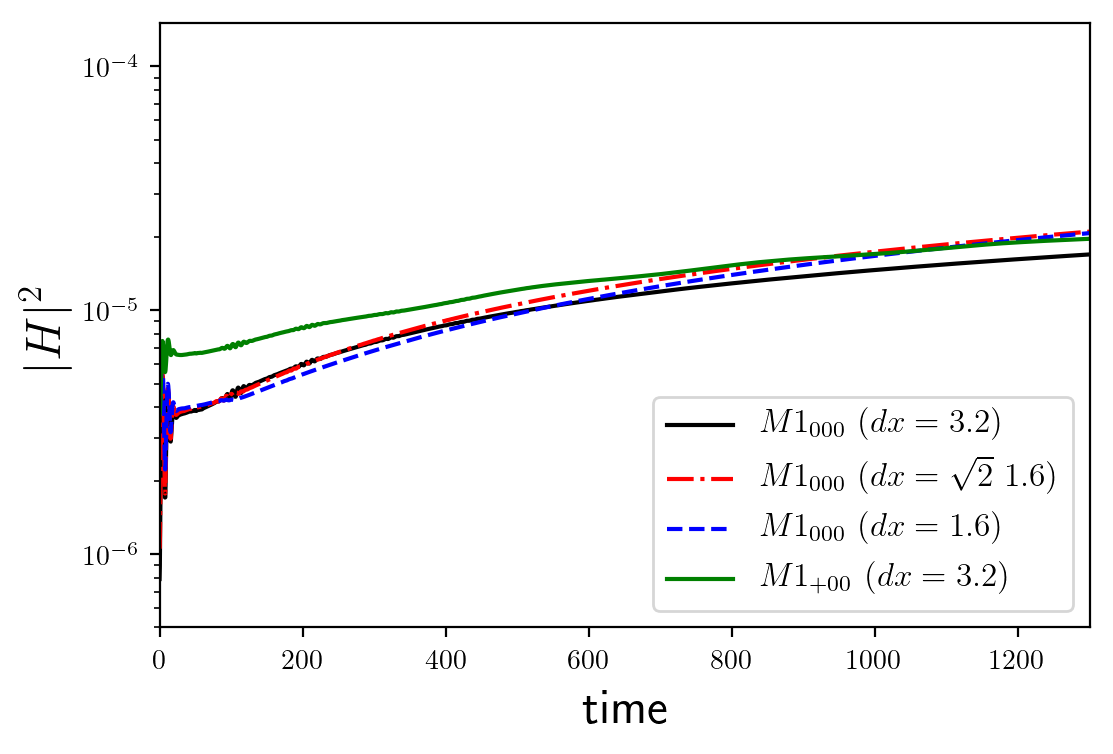}
\caption{Convergence for run $M1_{000}$: Evolution of the L2-norm of the Hamiltonian constraint for three different resolutions rescaled to show  fourth order convergence. Green line shows L2-norm of the Hamiltonian constrain for a perturbed run,
initially, the violation of the constraint due to the perturbation is evident, as time passes the magnitude of the error is comparable with the error of the unperturbed runs.}
 
\label{fig:H}
\end{figure}

The L2-norm of the Hamiltonian constrain, 
given by $|H|^2=\sqrt{\frac{\sum_{i=1}^N H_i^2}{N}}$, where $N$ is the number of points in the grid,
is shown in Fig. \ref{fig:H}. Here we conclude that the constraint equations converge as $H$ reduces when the resolution is increased, the black (solid) and the blue (dashed) line have been multiplied by the factors 4 and 16, showing fourth order convergence. The low resolution (solid line, coarsest grid $dx=3.2$) corresponds to the one used in all the simulations presented in this work.
The L2-norm of $H$ increases with time, however tends to a constant value which approaches zero as \{$dx, dy, dz\}\rightarrow 0$.

\begin{figure}
\includegraphics[width=0.5\textwidth]{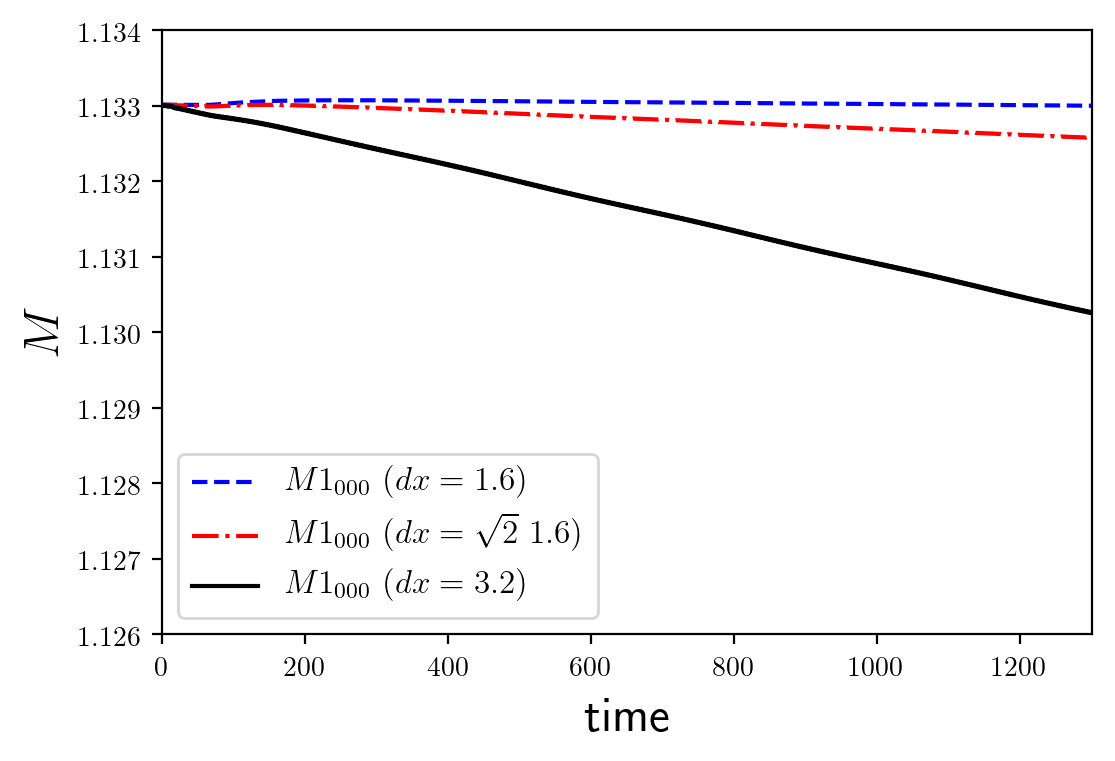}
\caption{Convergence for run $M1_{000}$: Evolution of the mass for three different resolutions.} 
\label{fig:Conv}
\end{figure}

We plot the mass for run $M1_{000}$. As the resolution is increased 
the mass converge to a constant value and the overall drift is reduced.
In Fig. \ref{fig:H} and Fig. \ref{fig:Conv} we have plotted until time equal to 1300, however the ($\{dx,dy,dz\}=3.2$) simulation extends up to $t\sim3500$, where the final total mass differs from the initial value by $0.6\%$.

\bibliographystyle{unsrt}
\bibliography{main} 
\end{document}